\documentclass[english]{article}
\usepackage[T1]{fontenc}
\usepackage[latin9]{inputenc}
\usepackage[a4paper]{geometry}
\usepackage{xcolor}
\usepackage{amsmath}
\usepackage{setspace}
\usepackage{esint}
\PassOptionsToPackage{normalem}{ulem}
\usepackage{ulem}
\makeatletter

\providecolor{lyxadded}{rgb}{0,0,1}
\providecolor{lyxdeleted}{rgb}{1,0,0}

\numberwithin{equation}{section}

\makeatother

\usepackage{babel}
\begin{document}
\begin{titlepage}

\global\long\def\thefootnote{\fnsymbol{footnote}}

~

\bigskip{}

\bigskip{}

\bigskip{}

\bigskip{}

\bigskip{}

\begin{center}
\textbf{\Large{}{}{}{}{}Super Yang--Mills action from WZW-like
open superstring field theory including the Ramond sector}{\Large{}{}{}{}
} 
\par\end{center}

\bigskip{}

\begin{center}
{\large{}{}{}{}{}{}{}{}}Mitsuru Asada$^1$ and Isao Kishimoto$^2${\large{}{}{}{}{}{}{}}
\par\end{center}{\large \par}

\begin{center}
$^1${\it Department of Physics, Niigata University, Niigata 950-2181,
Japan}\footnote{Present address: f15a001g@alumni.niigata-u.ac.jp}\\
$^2${\it Faculty of Education, Niigata University, Niigata 950-2181,
Japan}\footnote{E-mail: ikishimo@ed.niigata-u.ac.jp}
\par\end{center}

\bigskip{}

\bigskip{}

\bigskip{}

\bigskip{}

\bigskip{}

\bigskip{}

\bigskip{}

\begin{abstract}
In the framework of WZW-like open superstring field theory (SSFT)
including the Ramond (R) sector whose action was constructed by Kunitomo and Okawa,
we truncate the string fields in both the Neveu--Schwarz (NS)  and R sectors up to the lowest
massless level and obtain the ten-dimensional super Yang--Mills (SYM)
action with bosonic extra term by explicit calculation of the SSFT
action. Furthermore, we compute a contribution from the massive part
up to the lowest order and find that the bosonic extra term is canceled
and instead a fermionic extra term appears, which can be interpreted
as a string correction to the SYM action. This calculation is an extension
to the R sector of the earlier work by Berkovits and Schnabl in the NS
sector. We also study gauge transformation, equation of motion, and
spacetime supersymmetry transformation of the massless component fields
induced from those of string fields.
\end{abstract}
\global\long\def\thefootnote{\arabic{footnote}}

\end{titlepage}

\section{Introduction}

It is expected that open superstring theory describes super Yang--Mills
theory at low energy. Some time ago, Berkovits and Schnabl \cite{Berkovits:2003ny}
showed that the Yang--Mills action is derived from Berkovits' Wess--Zumino--Witten
(WZW)-like action for open superstring field theory \cite{Berkovits1995}
in the Neveu--Schwarz (NS) sector. Recently, Kunitomo and Okawa have
constructed a complete action for open superstring field theory including
the Ramond (R) sector as an extension of the WZW-like action \cite{Kunitomo:2015usa}.
Therefore, it is natural to expect that the ten-dimensional super
Yang--Mills (SYM) action can be derived from it. Furthermore, an explicit
formula for spacetime supersymmetry transformation in terms of string
fields has been proposed in Refs.~\cite{Kunitomo:2016kwh,Erler:2017onq}
and hence we can compare the induced transformation of the component
fields and the conventional supersymmetry transformation of SYM. 

As a first step toward the above issue, we adopt the level truncation
method at the lowest level, which corresponds to the massless component
fields due to the GSO projection. We perform explicit calculations
in the superstring field theory in both NS and R sectors and obtain the 
SYM action with extra bosonic term, $O(A_{\mu}^{4})$, in terms of
the component fields \cite{Asada:2017Th}. We compute a contribution
to the effective action of massless fields from the massive part of
string fields in the zero-momentum sector, up to lowest order with
respect to the coupling constant, and find that the extra $O(A_{\mu}^{4})$
is canceled\footnote{This result has already been obtained in Ref.~\cite{Berkovits:2003ny}
in computation in the NS sector.} and, instead, an extra fermionic term, $O(\lambda_{\dot{\alpha}}^{4})$,
appears in the order of $(\alpha^{\prime})^{1}$, which can be interpreted
as a string correction to SYM.

We also derive the induced gauge and supersymmetry transformations
up to nonzero lowest order and find that the resulting formulas are
consistent with those of the conventional SYM.

The organization of this paper is as follows. In the next section,
we briefly review the action of open superstring field theory in terms
of Kunitomo and Okawa's formulation. In Sect.~\ref{sec:Level-truncation},
we derive the explicit form of the level-truncated action at the lowest
level. In Sect.~\ref{sec:Contribution-from-massive}, we evaluate
the contribution from the massive part. In Sect.~\ref{sec:Induced-transformations},
we find the induced transformations from those of the superstring
field theory. In Sect.~\ref{sec:Concluding-remarks}, we give some
concluding remarks. In the appendix we summarize
our convention on spin fields, which is necessary for explicit computations.

\section{A brief review of a WZW-like action for open superstring field theory
including the R sector}

Let us first review Kunitomo and Okawa's action \cite{Kunitomo:2015usa}
for open superstring field theory (SSFT). 

As an extension of Berkovits' WZW-like action in the NS sector, Kunitomo
and Okawa proposed a complete action for open superstring field theory
including the R sector as follows:
\begin{align}
S[\Phi,\Psi] & =S_{\mathrm{NS}}[\Phi]+S_{\mathrm{R}}[\Phi,\Psi]\,,\label{eq:S_KO}\\
S_{\mathrm{NS}}[\Phi] & =-\int_{0}^{1}dt\langle A_{t}(t),QA_{\eta}(t)\rangle\,,\label{eq:S_NS}\\
S_{\mathrm{R}}[\Phi,\Psi] & =-\frac{1}{2}\langle\!\langle\Psi,YQ\Psi\rangle\!\rangle-\int_{0}^{1}dt\langle A_{t}(t),(F(t)\Psi)^{2}\rangle\,.\label{eq:S_R}
\end{align}
The action is a functional of string fields: $\Phi$ and $\Psi$.
$\langle\ ,\ \rangle$ and $\langle\!\langle\ ,\ \rangle\!\rangle$
are the BPZ inner product in the large and small Hilbert space, respectively.
$Q$ is the BRST operator and $\eta$ is the zeromode of $\eta(z)$
: $\eta=\oint\frac{dz}{2\pi i}\eta(z)$. We use the relation of the
superconformal ghosts between $(\beta,\gamma)$ and $(\xi,\eta,\phi)$
as $\beta(z)=\partial\xi e^{-\phi}(z)$ and  $\gamma(z)=e^{\phi}\eta(z)$.
$\Phi$ is a Grassmann even string field in the NS sector, whose ghost
number ($n_{\mathrm{gh}}$) and picture number ($n_{\mathrm{pic}}$)
are both zero, and expanded by the states in the large Hilbert space.
$A_{t}(t)$, $A_{\eta}(t)$, and $F(t)$ are defined by 
\begin{align}
A_{t}(t) & =(\partial_{t}e^{t\Phi})e^{-t\Phi}=\Phi\,,\qquad A_{\eta}(t)=(\eta e^{t\Phi})e^{-t\Phi}\,,\\
F(t)\Psi & =\Psi+\Xi\left\{ A_{\eta}(t),\Psi\right\} +\Xi\left\{ A_{\eta}(t),\Xi\left\{ A_{\eta}(t),\Psi\right\} \right\} +\cdots\\
 & =\sum_{k=0}^{\infty}\underbrace{\Xi\{A_{\eta}(t),\Xi\{A_{\eta}(t),\dots,\Xi\{A_{\eta}(t),}_{k}\Psi\}\dots\}\},
\end{align}
where $\Xi=\Theta(\beta_{0})$, the Heaviside step function of $\beta_{0}$.\footnote{In Ref.~\cite{Erler:2016ybs}, a better expression for $\Xi$ has been proposed.}
$\Psi$ is a Grassmann odd string field in the R sector with $(n_{\mathrm{gh}},n_{\mathrm{pic}})=(1,-1/2)$,
and it is expanded by the states in the \emph{restricted} small Hilbert
space. Namely, the conditions for the R string field are
\begin{equation}
\eta\Psi=0,\qquad XY\Psi=\Psi\,.
\end{equation}
$X$ and $Y$ are kinds of picture-changing operators with the picture
number $1$ and $-1$, respectively:
\begin{align}
X & =-\delta(\beta_{0})G_{0}+\delta^{\prime}(\beta_{0})b_{0}=\left\{ Q,\Theta(\beta_{0})\right\} \,,\\
Y & =-c_{0}\delta^{\prime}(\gamma_{0})\,,
\end{align}
where $XYX=X$ and $YXY=Y$ hold and therefore $XY$ gives a projector:
$(XY)^{2}=XY$.

In the following, we consider SSFT on the $N$ BPS D9-branes in the
flat ten-dimensional spacetime, and hence the string fields have the
Chan--Paton factors implicitly and the GSO projection is imposed on
the string fields.

\paragraph{Gauge transformation}

The action in Eq.~(\ref{eq:S_KO}) is invariant under the following infinitesimal
gauge transformations:
\begin{align}
A_{\delta_{g(\Lambda,\lambda,\Omega)}} & =Q\Lambda+D_{\eta}\Omega+\{F\Psi,F\Xi(\{F\Psi,\Lambda\}-\lambda)\}\,,\qquad A_{\delta}\equiv(\delta e^{\Phi})e^{-\Phi}\,,\label{eq:A_deltag}\\
\delta_{g(\Lambda,\lambda)}\Psi & =Q\lambda+X\eta F\Xi D_{\eta}(\{F\Psi,\Lambda\}-\lambda)\,,\label{eq:delta_gPsi}
\end{align}
where $F=F(t=1)$,
\begin{equation}
D_{\eta}B=\eta B-\left[A_{\eta},B\right\} \,,\qquad A_{\eta}=A_{\eta}(t=1)\,,
\end{equation}
 and the gauge parameter $\lambda$ in the R sector is in the restricted
small Hilbert space, namely,
\begin{equation}
\eta\lambda=0\,,\qquad XY\lambda=\lambda\,.
\end{equation}
$\Lambda$ and $\Omega$ are gauge parameters in the NS sector in
the large Hilbert space.

\paragraph{Equation of motion}

Taking the variation of the action in Eq.~(\ref{eq:S_KO}) with respect to
the string fields, $\Phi$ and $\Psi$, we have the equations of motion
as follows:
\begin{equation}
QA_{\eta}+(F\Psi)^{2}=0\,,\qquad Q\Psi+X\eta F\Psi=0\,.\label{eq:EOM}
\end{equation}
The second equation is consistent with the condition for $\Psi$,
which is in the restricted small Hilbert space.

\paragraph{Spacetime supersymmetry}

In the ten-dimensional Minkowski spacetime, the SSFT action in Eq.~(\ref{eq:S_KO})
is invariant under the spacetime supersymmetry transformation, which
is given by \cite{Kunitomo:2016kwh}:\footnote{Another form of supersymmetry transformation has been proposed in
Ref.~\cite{Erler:2017onq}. The difference does not matter in our paper
because we compute only the linearized one in Eq.~(\ref{eq:delta^0_S}).}
\begin{align}
A_{\delta_{\mathcal{S}}} & =e^{\Phi}\,\mathcal{S}\Xi(e^{-\Phi}F\Psi e^{\Phi})\,e^{-\Phi}+\{F\Psi,F\Xi A_{\mathcal{S}}\}\,,\qquad A_{\mathcal{S}}=(\mathcal{S}e^{\Phi})e^{-\Phi}\,,\label{eq:A_deltaS}\\
\delta_{\mathcal{S}}\Psi & =X\eta F\Xi\mathcal{S}A_{\eta}\,,\label{eq:delta_SPsi}
\end{align}
where $\mathcal{S}$ is a derivation with respect to the star product
of string fields and is given by a constant Weyl spinor $\epsilon_{\dot{\alpha}}$
and the spin field with $n_{\mathrm{pic}}=-1/2$ :
\begin{equation}
\mathcal{S}=\epsilon_{\dot{\alpha}}\oint\dfrac{dz}{2\pi i}S_{(-1/2)}^{\dot{\alpha}}(z)\,.
\end{equation}
We use a bosonized formulation of spin fields as in the appendix
for the following explicit computation. 

\section{Level truncation of string fields in the NS and R sectors\label{sec:Level-truncation}}

Here, we evaluate the SSFT action in Eq.~(\ref{eq:S_KO}) using the level
truncation method explicitly. We define the level of string fields
as the eigenvalue of $L_{0}=\left\{ Q,b_{0}\right\} $, except for a contribution
from momentum. We expand the string fields with component fields up
to the lowest level in both NS and R sectors.

\subsection{Level-truncated action in the NS sector}

In the NS sector, we expand a string field $\Phi$ with $(n_{\mathrm{gh}},n_{\mathrm{pic}})=(0,0)$
in the large Hilbert space. The lowest-level state is 
\begin{equation}
c\xi e^{-\phi}e^{ik\cdot X}(0)|0\rangle\,,
\end{equation}
which corresponds to the tachyon, but it is excluded by the GSO projection
of Eq.~(\ref{eq:GSOprojection}). The lowest-level states on the GSO projected
space are 
\begin{equation}
e^{ik\cdot X}(0)|0\rangle\,,\qquad c\xi\psi^{\mu}e^{-\phi}e^{ik\cdot X}(0)|0\rangle\,,\qquad c\partial c\xi\partial\xi e^{-2\phi}e^{ik\cdot X}(0)|0\rangle\,,
\end{equation}
which correspond to the massless level. The first one can be eliminated
by the $\Omega$-gauge transformation in Eq.~(\ref{eq:A_deltag}), which
can be rewritten as $e^{-\Phi}\delta_{g(\Omega)}e^{\Phi}=\eta(e^{-\Phi}\Omega e^{\Phi})$
and has an expression for finite transformation: $e^{\Phi^{\prime}}=e^{\Phi}g$
with $\eta g\thinspace=0$, because it can be rewritten as $e^{ik\cdot X}(0)|0\rangle=\eta\left(\xi e^{ik\cdot X}(0)|0\rangle\right)$.
In the following, we take into account only the other two states and
their component fields as a level-truncated string field $\Phi_{0}$
in the NS sector. Namely, we use a partial gauge-fixing condition,
$\xi_{0}\Phi=0$. After Ref.~\cite{Berkovits:2003ny}, we use the notation
\begin{align}
\Phi_{A} & =\int\frac{d^{10}k}{(2\pi)^{10}}A_{\mu}(k)\mathcal{V}_{A}^{\mu}(k)(0)|0\rangle\,,\ \ \ \ \mathcal{V}_{A}^{\mu}(k)=c\xi e^{-\phi}\psi^{\mu}(z)\,e^{ik\cdot X(z,\bar{z})}\,,\label{eq:V_A^mu}\\
\Phi_{B} & =\int\frac{d^{10}k}{(2\pi)^{10}}B(k)\mathcal{V}_{B}(k)(0)|0\rangle\,,\ \ \ \ \ \mathcal{V}_{B}(k)=c\partial c\xi\partial\xi e^{-2\phi}(z)\,e^{ik\cdot X(z,\bar{z})}\,,\label{eq:V_B}
\end{align}
where $A_{\mu}(k)$ and $B(k)$ are the Fourier modes of component
bosonic fields in the ten-dimensional spacetime. \\

The NS action in Eq.~(\ref{eq:S_NS}), which is the same as Berkovits' WZW-like
action, can be expanded as
\begin{align}
S_{\mathrm{NS}}[\Phi] & =\sum_{M,N=0}^{\infty}\frac{(-1)^{N}(M+N)!}{(M+N+2)!M!N!}\langle Q\Phi,\Phi^{M}(\eta\Phi)\Phi^{N}\rangle\nonumber \\
 & =\frac{1}{2}\langle Q\Phi,\eta\Phi\rangle+\frac{1}{3!}(\langle Q\Phi,\Phi\,\eta\Phi\rangle-\langle Q\Phi,\eta\Phi\,\Phi\rangle)\nonumber \\
 & \ \ +\frac{1}{4!}(\langle Q\Phi,\Phi^{2}\,\eta\Phi\rangle-2\langle Q\Phi,\Phi\,\eta\Phi\,\Phi\rangle+\langle Q\Phi,\eta\Phi\,\Phi^{2}\rangle)+O(\Phi^{5})\,,\label{eq:S_NS_expansion}
\end{align}
and we truncate the string field $\Phi$ in the NS sector to the sum
of Eqs.~(\ref{eq:V_A^mu})~and~(\ref{eq:V_B}): $\Phi_{0}=\Phi_{A}+\Phi_{B}$. 

Using the BRST transformations
\begin{align}
[Q,\mathcal{V}_{A}^{\mu}(k)] & =-\alpha^{\prime}k^{2}c\partial c\xi e^{-\phi}\psi^{\mu}e^{ik\cdot X}\nonumber \\
 & \quad-\sqrt{2\alpha^{\prime}}c(:k\cdot\psi\psi^{\mu}:+\frac{i}{\alpha^{\prime}}\partial X^{\mu}+(:\eta\xi:+\partial\phi)k^{\mu})e^{ik\cdot X}+\eta e^{\phi}\psi^{\mu}e^{ik\cdot X}\,,\\{}
[Q,\mathcal{V}_{B}(k)] & =-\sqrt{2\alpha^{\prime}}c\partial c\xi k\cdot\psi e^{-\phi}e^{ik\cdot X}+(-\partial c+2c(:\eta\xi:+\partial\phi))e^{ik\cdot X}
\end{align}
on the real axis, where $Q=\oint\frac{dz}{2\pi i}(c(T^{\mathrm{m}}+T^{\phi}+T^{\xi\eta})+bc\partial c+e^{\phi}\eta G^{\mathrm{m}}-\eta\partial\eta e^{2\phi}b)$,
$T^{\mathrm{m}}=-\frac{1}{\alpha'}\partial X^{\mu}\partial X_{\mu}-\frac{1}{2}\psi^{\mu}\partial\psi_{\mu}$,
and $G^{\mathrm{m}}=i\sqrt{\frac{2}{\alpha'}}\psi^{\mu}\partial X_{\mu}$,
we can evaluate the kinetic term $\frac{1}{2}\langle Q\Phi_{0},\eta\Phi_{0}\rangle$.
For the interaction terms, we use the explicit form of the conformal
maps, which defines the $n$-string term. Namely, for $A_{k}=A_{k}(0)|0\rangle$
($k=1,2,\dots,n$), we have
\begin{align}
\langle A_{1},\,A_{2}\cdots A_{n}\rangle & =\langle g_{1}^{(n)}\circ A_{1}(0)\,g_{2}^{(n)}\circ A_{2}(0)\,\cdots\,g_{n}^{(n)}\circ A_{n}(0)\rangle_{\mathrm{UHP}}\,,\\
g_{k}^{(n)}(z) & =h^{-1}(e^{i\pi\frac{2k-1-n}{n}}(h(z))^{\frac{2}{n}})=\tan\left(\frac{2}{n}\arctan z+\frac{\pi}{2n}(2k-1-n)\right)\,,
\end{align}
with the map from the upper half-plane to the unit disk: $h(z)=\frac{1+iz}{1-iz}$.
The normalization of the large Hilbert space is given by
\begin{equation}
\langle\xi(y)\frac{1}{2}c\partial c\partial^{2}c(z)e^{-2\phi(w)}e^{ik\cdot X(x,\bar{x})}\rangle_{\mathrm{UHP}}=(2\pi)^{10}\delta^{10}(k)\,.\label{eq:normalization_large}
\end{equation}
In particular, with respect to the $\phi$-charge, the terms such
as $\Phi^{n-2}\eta\Phi\,Q\Phi$ are linear combinations of $e^{q\phi}$
with $q\le2-n$ because $\Phi_{A}\sim e^{-\phi};\,\eta\Phi_{A}\sim e^{-\phi};\,\Phi_{B}\sim e^{-2\phi};\,\eta\Phi_{B}\sim e^{-2\phi};\,Q\Phi_{A}\sim e^{-\phi},1,e^{\phi}$
and $Q\Phi_{B}\sim e^{-\phi},1$, which imply that the higher-order
interaction terms, $O(\Phi^{5})$ in Eq.~(\ref{eq:S_NS_expansion}), vanish
for $\Phi_{0}=\Phi_{A}+\Phi_{B}$. Then, we obtain
\begin{align}
 & S_{\mathrm{NS}}[\Phi_{0}]=\int d^{10}x\,\mathrm{Tr}\left[\frac{\alpha^{\prime}}{2}A_{\mu}\partial^{2}A^{\mu}+i\sqrt{2\alpha^{\prime}}B\partial^{\mu}A_{\mu}+B^{2}+\frac{i\sqrt{2\alpha^{\prime}}}{2}\partial_{\mu}\tilde{A}_{\nu}[\tilde{A}^{\mu},\tilde{A}^{\nu}]\right]\label{eq:S_NSPhi_0}\\
 & +\!\int\!d^{10}x2^{\frac{\alpha^{\prime}}{2}\left((\partial_{1}-\partial_{3})^{2}+(\partial_{2}-\partial_{4})^{2}\right)}\mathrm{Tr}\!\left[\frac{1}{8}A_{\mu}(x_{1})A_{\nu}(x_{2})A^{\mu}(x_{3})A^{\nu}(x_{4})\!-\!\frac{1}{2}A_{\mu}(x_{1})A_{\nu}(x_{2})A^{\nu}(x_{3})A^{\mu}(x_{4})\!\right]_{x_{i}=x}\!,\nonumber 
\end{align}
where the trace is taken over the indices of the Chan--Paton factors,
$T_{a}$; $A_{\mu}(x)$ and $B(x)$ are given by
\begin{equation}
A_{\mu}(x)=A_{\mu}^{a}(x)T_{a}=\int\frac{d^{10}k}{(2\pi)^{10}}A_{\dot{\mu}}(k)e^{ik\cdot x}\,,\qquad B(x)=B^{a}(x)T_{a}=\int\frac{d^{10}k}{(2\pi)^{10}}B(k)e^{ik\cdot x}\,;
\end{equation}
 and $\tilde{A}_{\mu}$ is defined as
\begin{equation}
\tilde{A}_{\mu}(x)=K^{-\alpha^{\prime}\partial^{2}}A_{\mu}(x)\,,\qquad K\equiv\frac{4}{3\sqrt{3}}\,.\label{eq:tileA_mu}
\end{equation}
We note that the reality condition for the NS string field, $\mathrm{bpz^{-1}}\circ\Phi^{\dagger}=-\Phi$
\cite{Ohmori:2003vq}, implies the (anti-)Hermiticity of the component
fields: $A_{\mu}(x)^{\dagger}=A_{\mu}(x)$ and $B(x)^{\dagger}=-B(x)$.

Integrating out the scalar component field $B(x)$ in Eq.~(\ref{eq:S_NSPhi_0}),\footnote{We note that $\Phi_{B}$ does not contribute to the terms including
the R string fields, up to the lowest level, as we see in Sect.~\ref{subsec:Level-truncated-R}.} or using the equation of motion for $B$, $B=-i\sqrt{\alpha^{\prime}/2}\,\partial^{\mu}A_{\mu}$,
and taking the small-momentum limit: $K^{-\alpha^{\prime}\partial^{2}}\sim 1$
and $2^{\frac{\alpha^{\prime}}{2}\left((\partial_{1}-\partial_{3})^{2}+(\partial_{2}-\partial_{4})^{2}\right)}\sim 1$,
we have
\begin{align}
S_{\mathrm{NS}}[\Phi_{0}] & =-\frac{\alpha^{\prime}}{4}\int d^{10}x\,\mathrm{Tr}\left(\partial_{\mu}A_{\nu}-\partial_{\nu}A_{\mu}-\frac{i}{\sqrt{2\alpha^{\prime}}}[A_{\mu,}A_{\nu}]\right)^{2}\nonumber \\
 & \quad-\frac{1}{4}\int d^{10}x\,\mathrm{Tr}[\frac{1}{2}A_{\mu}A_{\nu}A^{\mu}A^{\nu}+A_{\mu}A_{\nu}A^{\nu}A^{\mu}]\,.\label{eq:S_NS_A4}
\end{align}
The first line on the right-hand side corresponds to the ordinary
Yang--Mills action and the second line is the difference from it. Actually,
it is known that the difference is canceled by the contribution from
massive component fields \cite{Berkovits:2003ny}. 

\subsection{Level-truncated action including the R string field\label{subsec:Level-truncated-R}}

In the R sector, the string field $\Psi$, which has $(n_{\mathrm{gh}},n_{\mathrm{pic}})=(1,-1/2)$,
is in the restricted small Hilbert space. The lowest-level states
are given by spin fields with $(-1/2)$-picture:
\begin{equation}
cS_{(-1/2)}^{\alpha}e^{ik\cdot X}(0)|0\rangle\,,\qquad cS_{(-1/2)}^{\dot{\alpha}}e^{ik\cdot X}(0)|0\rangle\,,
\end{equation}
where $\alpha$ and $\dot{\alpha}$ are spinor indices with $16$
components, and they correspond to the massless level. By the GSO
projection in Eq.~(\ref{eq:GSOprojection}), only the state with the dotted spinor
remains in our convention. Hence, we expand the level-truncated
string field $\Psi_{0}$ as
\begin{equation}
\Psi_{0}=\int\frac{d^{10}k}{(2\pi)^{10}}\lambda_{\dot{\alpha}}(k)\mathcal{V}_{\lambda}^{\dot{\alpha}}(k)(0)|0\rangle\,,\ \ \ \ \mathcal{V}_{\lambda}^{\dot{\alpha}}(k)=cS_{(-1/2)}^{\dot{\alpha}}(z)\,e^{ik\cdot X(z,\bar{z})}\,,\label{eq:Psi_0_def}
\end{equation}
where $\lambda_{\dot{\alpha}}(k)$ is the Fourier mode of a fermionic
component field, which is the Weyl spinor in the ten-dimensional spacetime.
This string field $\Psi_{0}$ is Grassmann odd, and we find that it
is indeed in the restricted small Hilbert space: $\eta\Psi_{0}=0$
and $XY\Psi_{0}=\Psi_{0}$.\\

Similarly to the NS sector, we derive an explicit expression of the
action including the R string field in terms of the component fields.
The action in Eq.~(\ref{eq:S_R}) is expanded as
\begin{equation}
S_{\mathrm{R}}[\Phi,\Psi]=-\frac{1}{2}\langle\!\langle\Psi,YQ\Psi\rangle\!\rangle-\langle\Phi,\Psi^{2}\rangle+O(\Phi^{2}\Psi^{2})\,,\label{eq:S_Rexpand}
\end{equation}
and we truncate the string fields up to the lowest level: the NS string
field $\Phi$ to $\Phi_{0}=\Phi_{A}+\Phi_{B}$ and the R string field
$\Psi$ to $\Psi_{0}$ given in Eq.~(\ref{eq:Psi_0_def}).

The kinetic term in the R sector is evaluated using the normalization
of the small Hilbert space:
\begin{equation}
\langle\!\langle\frac{1}{2}c\partial c\partial^{2}c(z)e^{-2\phi(w)}e^{ik\cdot X(x,\bar{x})}\rangle\!\rangle_{\mathrm{UHP}}=(2\pi)^{10}\delta^{10}(k)\,.
\end{equation}
It is convenient to use $\langle\!\langle\Psi,YQ\Psi\rangle\!\rangle=\langle\!\langle\Psi,Y_{\mathrm{mid}}Q\Psi\rangle\!\rangle$
\cite{Kunitomo:2015usa} for calculation, where $Y_{\mathrm{mid}}=Y(i)$
is the midpoint insertion of the conventional inverse picture-changing
operator $Y(z)=c\partial\xi e^{-2\phi}(z)$. With the BRST transformation
\begin{equation}
[Q,\mathcal{V}_{\lambda}^{\dot{\alpha}}(k)]=-\alpha^{\prime}k^{2}c\partial cS_{(-1/2)}^{\dot{\alpha}}e^{ik\cdot X}-i\sqrt{\alpha^{\prime}}k_{\mu}(\Gamma^{\mu})_{\ \beta}^{\dot{\alpha}}\eta cS_{(1/2)}^{\beta}e^{ik\cdot X}
\end{equation}
on the real axis, we have
\begin{align}
-\frac{1}{2}\langle\!\langle\Psi_{0},Y_{\mathrm{mid}}Q\Psi_{0}\rangle\!\rangle & =\frac{\sqrt{\alpha^{\prime}}}{2}\int d^{10}x\,\mathrm{Tr}\left[\lambda_{\dot{\alpha}}(\Gamma^{\mu}C)^{\dot{\alpha}\dot{\beta}}\partial_{\mu}\lambda_{\dot{\beta}}\right]\,,\label{eq:Psi_0YmidQPsi_0}
\end{align}
where
\begin{equation}
\lambda_{\dot{\alpha}}(x)=\lambda_{\dot{\alpha}}^{a}(x)T_{a}=\int\frac{d^{10}k}{(2\pi)^{10}}\lambda_{\dot{\alpha}}(k)e^{ik\cdot x}\,.
\end{equation}
For the cubic interaction term, we find $\langle\Phi_{B},\Psi_{0}^{2}\rangle=0$
by counting the number of $c$ ghosts. With a similar manipulation
to the NS sector, the remaining term is evaluated as
\begin{equation}
-\langle\Phi_{A},\Psi_{0}^{2}\rangle=\frac{i}{\sqrt{2}}\int d^{10}x\,\mathrm{Tr}\left[\tilde{A}_{\mu}\tilde{\lambda}_{\dot{\alpha}}(\Gamma^{\mu}C)^{\dot{\alpha}\dot{\beta}}\tilde{\lambda}_{\dot{\beta}}\right]\,,\label{eq:Psi_0cubic}
\end{equation}
where
\begin{equation}
\tilde{\lambda}_{\dot{\alpha}}(x)=K^{-\alpha^{\prime}\partial^{2}}\lambda_{\dot{\alpha}}(x)
\end{equation}
as in Eq.~(\ref{eq:tileA_mu}). For higher interaction terms, we find
that $O(\Phi^{2}\Psi^{2})$ in Eq.~(\ref{eq:S_Rexpand}) with $\Phi=\Phi_{0}$
and $\Psi=\Psi_{0}$ vanishes thanks to the normalization in Eq.~(\ref{eq:normalization_large})
because $\Phi_{A}\sim c,\,\Phi_{B}\sim cc$, and $\Psi_{0}\sim c$,
with respect to the $bc$-ghost sector. From Eqs.~(\ref{eq:Psi_0YmidQPsi_0}),
(\ref{eq:Psi_0cubic}), and (\ref{eq:S_Rexpand}), for the small-momentum
limit $K^{-\alpha^{\prime}\partial^{2}}\sim1$, we have obtained

\begin{equation}
S_{\mathrm{R}}[\Phi_{0},\Psi_{0}]=-\frac{\sqrt{\alpha^{\prime}}}{2}\int d^{10}x\,\mathrm{Tr}\left[i\hat{\lambda}^{\alpha}(C\Gamma^{\mu})_{\alpha\beta}D_{\mu}\hat{\lambda}^{\beta}\right],\label{eq:S_R_massless}
\end{equation}
where 
\begin{align}
\hat{\lambda}^{\alpha} & =C^{\alpha\dot{\beta}}\lambda_{\dot{\beta}}\,,\label{eq:hat_lambda_def}\\
D_{\mu}\hat{\lambda} & =\partial_{\mu}\hat{\lambda}-\frac{i}{\sqrt{2\alpha^{\prime}}}[A_{\mu},\hat{\lambda}]\,.
\end{align}
The above form of $S_{\mathrm{R}}[\Phi_{0},\Psi_{0}]$ just corresponds
to the gaugino term of the ten-dimensional SYM action.

\section{Contribution from the massive part\label{sec:Contribution-from-massive}}

Here, we consider a contribution to the effective action of massless
fields, $A_{\mu}$ and $\hat{\lambda}^{\alpha}$, obtained in the
previous section, from the massive part of string fields, $\Phi$ and
$\Psi$, in the same way as Ref.~\cite{Berkovits:2003ny}. The SSFT action
with the coupling constant $g$, which is obtained by replacing $g^{-2}S[g\Phi,g\Psi${]}
using Eq.~(\ref{eq:S_KO}) with $S[\Phi,\Psi]$, can be expanded as
\begin{align}
S[\Phi,\Psi] & =-\frac{1}{2}\langle\eta\Phi,Q\Phi\rangle+\frac{g}{6}\langle\eta\Phi,[\Phi,Q\Phi]\rangle\nonumber \\
 & \quad-\frac{1}{2}\langle\!\langle\Psi,YQ\Psi\rangle\!\rangle-g\langle\Phi,\Psi^{2}\rangle+O(g^{2})\,.\label{eq:S=00005BPhiPsi=00005D__g}
\end{align}
In this section, we concentrate on the zero-momentum
sector, and then the level-truncated string fields, $\Phi_{0},\,\Psi_{0}$,
satisfy $Q\eta\Phi_{0}=0,\ Q\Psi_{0}=0$ because of  Eqs.~(\ref{eq:EOM_Phi_0})~and~(\ref{eq:EOM_Psi_0}) with $B=-i\sqrt{\alpha^{\prime}/2}\,\partial^{\mu}A_{\mu}$.
We also note that $\Phi_{0}$ satisfies $\xi_{0}\eta\Phi_{0}=\Phi_{0}$,
namely, the partial gauge-fixing condition. Around the massless part
of the string fields, $\Phi_{0},\,\Psi_{0}$, we expand the string fields
$\Phi,\,\Psi$ as $\Phi_{0}+R,\,\Psi_{0}+S$, where $R$ and $S$
are the massive part in the NS and R sector respectively, and we have
\begin{align}
S[\Phi_{0}+R,\Psi_{0}+S]-S[\Phi_{0},\Psi_{0}] & =-\frac{1}{2}\langle\eta R,QR\rangle+\frac{g}{2}\langle\eta R,[\Phi_{0},Q\Phi_{0}]\rangle-g\langle R,(\Psi_{0})^{2}\rangle\nonumber \\
 & \quad-\frac{1}{2}\langle\!\langle S,YQS\rangle\!\rangle-g\langle\!\langle S,\left\{ \eta\Phi_{0},\Psi_{0}\right\} \rangle\!\rangle+\cdots\label{eq:S__RSexpand}
\end{align}
from Eq.~(\ref{eq:S=00005BPhiPsi=00005D__g}). Here, $(+\cdots)$ denotes
the higher-order terms in $g$ when $R$ and $S$ are assumed to be
$O(g)$. Varying the above with respect to the massive part, we have
equations of motion
\begin{align}
 & Q\eta R=g\left(-\frac{1}{2}\left\{ \eta\Phi_{0},Q\Phi_{0}\right\} -(\Psi_{0})^{2}\right)+\cdots\,,\qquad QS=g\left(-X\left\{ \eta\Phi_{0},\Psi_{0}\right\} \right)+\cdots\,,\label{eq:EOM_RS}
\end{align}
and these can be solved by using the propagators in Ref.~\cite{Kunitomo:2016bhc}
as
\begin{equation}
R_{{\rm s}}=-g\frac{\xi_{0}b_{0}}{L_{0}}\left(\frac{1}{2}\eta[\Phi_{0},Q\Phi_{0}]+(\Psi_{0})^{2}\right),\qquad S_{{\rm s}}=-g\frac{b_{0}X\eta}{L_{0}}\left[\Phi_{0},\Psi_{0}\right]\,,\label{eq:R_s-S_s}
\end{equation}
up to lowest order in $g$, where $R_{{\rm s}}$ satisfies the partial
gauge-fixing condition: $\xi_{0}R_{{\rm s}}=0$.

Expanding the massive part of string fields around Eq.~(\ref{eq:R_s-S_s})
as $R=R^{\prime}+R_{{\rm s}},\ S=S^{\prime}+S_{{\rm s}}$, we
have
\begin{align}
 & S[\Phi_{0}+R^{\prime}+R_{{\rm s}},\Psi_{0}+S^{\prime}+S_{{\rm s}}]-S[\Phi_{0},\Psi_{0}]\nonumber \\
 & =\frac{1}{2}\langle QR^{\prime},\eta R^{\prime}\rangle-\frac{1}{2}\langle\!\langle S^{\prime},YQS^{\prime}\rangle\!\rangle\nonumber \\
 & \quad-\frac{1}{2}\langle QR_{{\rm s}},\eta R_{{\rm s}}\rangle+\frac{1}{2}\langle\!\langle S_{{\rm s}},YQS_{{\rm s}}\rangle\!\rangle+\cdots,\label{eq:SR'S'expand}
\end{align}
where linear terms with respect to $R^{\prime}$ and $S^{\prime}$
vanish and therefore the massive part decouples from the massless
part. The second line on the right-hand side gives a contribution
to the effective action of the massless fields, which can be computed
as
\begin{align}
-\frac{1}{2}\langle QR_{{\rm s}},\eta R_{{\rm s}}\rangle+\frac{1}{2}\langle\!\langle S_{{\rm s}},YQS_{{\rm s}}\rangle\!\rangle & =\frac{g^{2}}{8}\langle[\Phi_{0},Q\Phi_{0}],\frac{b_{0}}{L_{0}}\left\{ \eta\Phi_{0},Q\Phi_{0}\right\} \rangle+\frac{g^{2}}{2}\langle\!\langle(\Psi_{0})^{2},\frac{b_{0}}{L_{0}}(\Psi_{0})^{2}\rangle\!\rangle\nonumber \\
 & \quad-\frac{g^{2}}{2}\langle(\Psi_{0})^{2},\frac{b_{0}}{L_{0}}[\Phi_{0},Q\Phi_{0}]\rangle-\frac{g^{2}}{2}\langle\left\{ Q\Phi_{0},\Psi_{0}\right\} ,\frac{b_{0}}{L_{0}}[\Phi_{0},\Psi_{0}]\rangle\,.\label{eq:effective}
\end{align}
In the computation for the term including $S_{{\rm s}}$, we manipulated
$X$ as in Ref.~\cite{Kunitomo:2016bhc}. Namely, we rewrote it as
\begin{align}
\frac{1}{2}\langle\!\langle S_{{\rm s}},YQS_{{\rm s}}\rangle\!\rangle & =\frac{g^{2}}{2}\langle\!\langle\left\{ \eta\Phi_{0},\Psi_{0}\right\} ,\frac{b_{0}X}{L_{0}}\left\{ \eta\Phi_{0},\Psi_{0}\right\} \rangle\!\rangle\nonumber \\
 & =\frac{g^{2}}{2}\langle\xi_{0}\eta\Phi_{0},\Psi_{0}\,\frac{b_{0}X}{L_{0}}\left\{ \eta\Phi_{0},\Psi_{0}\right\} \rangle+\frac{g^{2}}{2}\langle\xi_{0}\eta\Phi_{0},\,\frac{b_{0}X}{L_{0}}\left\{ \eta\Phi_{0},\Psi_{0}\right\} \Psi_{0}\rangle\nonumber \\
 & =\frac{g^{2}}{2}\langle[\Phi_{0},\Psi_{0}],\frac{b_{0}X}{L_{0}}\left\{ \eta\Phi_{0},\Psi_{0}\right\} \rangle\nonumber \\
 & =\frac{g^{2}}{2}\langle[\Phi_{0},\Psi_{0}],\frac{b_{0}}{L_{0}}Q\Xi\left\{ \eta\Phi_{0},\Psi_{0}\right\} \rangle=\frac{g^{2}}{2}\langle[\Phi_{0},\Psi_{0}],\left(1-Q\frac{b_{0}}{L_{0}}\right)\Xi\left\{ \eta\Phi_{0},\Psi_{0}\right\} \rangle\nonumber \\
 & =\frac{g^{2}}{2}\langle[\Phi_{0},\Psi_{0}],\Xi\left\{ \eta\Phi_{0},\Psi_{0}\right\} \rangle-\frac{g^{2}}{2}\langle\left\{ Q\Phi_{0},\Psi_{0}\right\} ,\frac{b_{0}}{L_{0}}[\Phi_{0},\Psi_{0}]\rangle\,,
\end{align}
where the first term of the last expression vanishes due to the number
of $c$-ghosts and the normalization in Eq.~(\ref{eq:normalization_large}).

The first term of the right-hand side of Eq.~(\ref{eq:effective}) has
been evaluated in Ref.~\cite{Berkovits:2003ny}:
\begin{align}
 & \frac{1}{8}\langle[\Phi_{0},Q\Phi_{0}],\frac{b_{0}}{L_{0}}\left\{ \eta\Phi_{0},Q\Phi_{0}\right\} \rangle\nonumber \\
 & =-\int d^{10}x{\rm Tr}[A_{\mu}A_{\nu}A_{\rho}A_{\sigma}]\int_{0}^{\infty}dt\,e^{-t}(e^{-2t}a^{2}-\frac{1}{a^{2}})\left(\frac{\eta^{\mu\rho}\eta^{\nu\sigma}}{(a^{-1}+e^{-t}a)^{4}}+\frac{\eta^{\mu\sigma}\eta^{\nu\rho}}{(a^{-1}-e^{-t}a)^{4}}\right)\nonumber \\
 & =\frac{1}{4}\int d^{10}x\,\mathrm{Tr}[\frac{1}{2}A_{\mu}A_{\nu}A^{\mu}A^{\nu}+A_{\mu}A_{\nu}A^{\nu}A^{\mu}]\,,\label{eq:YMcancel}
\end{align}
where 
\begin{equation}
a=\tan\frac{\pi}{8}=\sqrt{2}-1\,.\label{eq:a_def}
\end{equation}
Equation (\ref{eq:YMcancel}) cancels the extra term of Eq.~(\ref{eq:S_NS_A4})
in the NS sector. We note that the component field $B$ in $\Phi_{0}$
does not contribute to the above thanks to the equation of motion,
$B=-i\sqrt{\alpha^{\prime}/2}\,\partial^{\mu}A_{\mu}$ in the lowest
order in $g$, which is negligible in the zero-momentum sector. The
remaining terms of the right-hand side of Eq.~(\ref{eq:effective}), which
includes the R sector, do not shift the coefficient of the $O(A_{\mu}\hat{\lambda}\hat{\lambda})$
term of Eq.~(\ref{eq:S_R_massless}). Furthermore, we can neglect the
third and fourth terms in Eq.~(\ref{eq:effective}), namely
\begin{equation}
-\frac{1}{2}\langle(\Psi_{0})^{2},\frac{b_{0}}{L_{0}}[\Phi_{0},Q\Phi_{0}]\rangle=0,\qquad-\frac{1}{2}\langle\left\{ Q\Phi_{0},\Psi_{0}\right\} ,\frac{b_{0}}{L_{0}}[\Phi_{0},\Psi_{0}]\rangle=0\,,
\end{equation}
because of Eq.~(\ref{eq:normalization_large}), noting that $\mathcal{V}_{A}^{\mu}(k=0)=c\xi e^{-\phi}\psi^{\mu},\ [Q,\mathcal{V}_{A}^{\mu}(k=0)]=-i\sqrt{2/\alpha^{\prime}}\,c\partial X^{\mu}+\eta e^{\phi}\psi^{\mu},\ \mathcal{V}_{\lambda}^{\dot{\alpha}}(k=0)=cS_{(-1/2)}^{\dot{\alpha}}$.
The second term of Eq.~(\ref{eq:effective}) can be evaluated, in a similar
way to Eq.~(\ref{eq:YMcancel}), as
\begin{align}
 & \frac{1}{2}\langle\!\langle(\Psi_{0})^{2},\frac{b_{0}}{L_{0}}(\Psi_{0})^{2}\rangle\!\rangle=\frac{1}{2}\int d^{10}x{\rm Tr}[\lambda_{\dot{\alpha}}\lambda_{\dot{\beta}}\lambda_{\dot{\gamma}}\lambda_{\dot{\delta}}]\langle\!\langle cS_{(-1/2)}^{\dot{\alpha}}\ast cS_{(-1/2)}^{\dot{\beta}},\frac{b_{0}}{L_{0}}(cS_{(-1/2)}^{\dot{\gamma}}\ast cS_{(-1/2)}^{\dot{\delta}})\rangle\!\rangle\nonumber \\
 & =\frac{1}{2}\int d^{10}x{\rm Tr}[\lambda_{\dot{\alpha}}\lambda_{\dot{\beta}}\lambda_{\dot{\gamma}}\lambda_{\dot{\delta}}]\langle0|cS_{(-1/2)}^{\dot{\alpha}}(-\sqrt{3})cS_{(-1/2)}^{\dot{\beta}}(\sqrt{3})U_{3}\frac{b_{0}}{L_{0}}U_{3}^{\dagger}cS_{(-1/2)}^{\dot{\gamma}}(\frac{1}{\sqrt{3}})cS_{(-1/2)}^{\dot{\delta}}(\frac{-1}{\sqrt{3}})|0\rangle\nonumber \\
 & =\frac{1}{2}\int d^{10}x{\rm Tr}[\lambda_{\dot{\alpha}}\lambda_{\dot{\beta}}\lambda_{\dot{\gamma}}\lambda_{\dot{\delta}}]\int_{0}^{\infty}dt\langle S_{(-1/2)}^{\dot{\alpha}}(-\frac{1}{a})S_{(-1/2)}^{\dot{\beta}}(\frac{1}{a})S_{(-1/2)}^{\dot{\gamma}}(e^{-t}a)S_{(-1/2)}^{\dot{\delta}}(-e^{-t}a)\rangle\nonumber \\
 & \qquad\times\langle0|c(-\frac{1}{a})c(\frac{1}{a})b_{0}c(e^{-t}a)c(-e^{-t}a)|0\rangle\,,\label{eq:Psi0^4_org}
\end{align}
where $a$ is given in Eq.~(\ref{eq:a_def}) and $U_{3}$ is given in Refs.~\cite{Schnabl2003,Schnabl2006},
which corresponds to the conformal map $\tan\left(\frac{2}{3}\arctan z\right)$.
For the $bc$-ghost sector, we have
\begin{equation}
\langle0|c(-\frac{1}{a})c(\frac{1}{a})b_{0}c(e^{-t}a)c(-e^{-t}a)|0\rangle=-4e^{-t}(e^{-2t}a^{2}-\frac{1}{a^{2}})\label{eq:bc_a_result}
\end{equation}
with the normalization $\langle0|c_{-1}c_{0}c_{1}|0\rangle=1$. As
for the $\phi$-ghost and the spin field sector, we note that\footnote{The last expression is consistent with the formula in Ref.~\cite{Kostelecky:1986xg}
using the Fierz identity:
\begin{equation}
(\Gamma^{\mu}C)^{\dot{\alpha}\dot{\beta}}(\Gamma_{\mu}C)^{\dot{\gamma}\dot{\delta}}+(\Gamma^{\mu}C)^{\dot{\alpha}\dot{\gamma}}(\Gamma_{\mu}C)^{\dot{\delta}\dot{\beta}}+(\Gamma^{\mu}C)^{\dot{\alpha}\dot{\delta}}(\Gamma_{\mu}C)^{\dot{\beta}\dot{\gamma}}=0\,.
\end{equation}
}
\begin{align}
 & \langle S_{(-1/2)}^{\dot{\alpha}_{1}}(z_{1})S_{(-1/2)}^{\dot{\alpha}_{2}}(z_{2})S_{(-1/2)}^{\dot{\alpha}_{3}}(z_{3})S_{(-1/2)}^{\dot{\alpha}_{4}}(z_{4})\rangle\nonumber \\
 & =\delta_{\dot{A}_{1}+\dot{A}_{2}+\dot{A}_{3}+\dot{A}_{4},0}\left(\prod_{p<q}(z_{pq})^{\sum_{i=1}^{5}\dot{A}_{p}^{i}\dot{A}_{q}^{i}-\frac{1}{4}}\right)\exp\left[i\pi\left(\sum_{p<q}\sum_{i,j=1}^{5}\dot{A}_{p}^{i}M_{ij}\dot{A}_{q}^{j}+\frac{1}{2}\sum_{j=1}^{5}M_{6j}(\dot{A}_{2}^{j}+\dot{A}_{4}^{j})\right)\right]\nonumber \\
 & =\frac{1}{2z_{12}z_{13}z_{24}z_{34}}(\Gamma^{\mu}C)^{\dot{\alpha}_{1}\dot{\alpha}_{2}}(\Gamma_{\mu}C)^{\dot{\alpha}_{3}\dot{\alpha}_{4}}-\frac{1}{2z_{13}z_{14}z_{23}z_{24}}(\Gamma^{\mu}C)^{\dot{\alpha}_{4}\dot{\alpha}_{1}}(\Gamma_{\mu}C)^{\dot{\alpha}_{2}\dot{\alpha}_{3}}\,,
\end{align}
where $z_{pq}=z_{p}-z_{q}$, and it leads to
\begin{align}
 & \langle S_{(-1/2)}^{\dot{\alpha}}(-\frac{1}{a})S_{(-1/2)}^{\dot{\beta}}(\frac{1}{a})S_{(-1/2)}^{\dot{\gamma}}(e^{-t}a)S_{(-1/2)}^{\dot{\delta}}(-e^{-t}a)\rangle\nonumber \\
 & =\frac{e^{t}}{8(a^{-1}+e^{-t}a)^{2}}(\Gamma^{\mu}C)^{\dot{\alpha}\dot{\beta}}(\Gamma_{\mu}C)^{\dot{\gamma}\dot{\delta}}-\frac{1}{2(a^{-2}-e^{-2t}a^{2})^{2}}(\Gamma^{\mu}C)^{\dot{\delta}\dot{\alpha}}(\Gamma_{\mu}C)^{\dot{\beta}\dot{\gamma}}\,.
\end{align}

With the above and Eq.~(\ref{eq:bc_a_result}), Eq.~(\ref{eq:Psi0^4_org})
is rewritten as
\begin{align}
 & \frac{1}{2}\langle\!\langle(\Psi_{0})^{2},\frac{b_{0}}{L_{0}}(\Psi_{0})^{2}\rangle\!\rangle\nonumber \\
 & =\frac{1}{4}\int d^{10}x{\rm Tr}[\lambda_{\dot{\alpha}}\lambda_{\dot{\beta}}\lambda_{\dot{\gamma}}\lambda_{\dot{\delta}}]\int_{0}^{\infty}dt\left(\frac{a^{-1}-e^{-t}a}{a^{-1}+e^{-t}a}(\Gamma^{\mu}C)^{\dot{\alpha}\dot{\beta}}(\Gamma_{\mu}C)^{\dot{\gamma}\dot{\delta}}-\frac{4e^{-t}}{a^{-2}-e^{-2t}a^{2}}(\Gamma^{\mu}C)^{\dot{\delta}\dot{\alpha}}(\Gamma_{\mu}C)^{\dot{\beta}\dot{\gamma}}\right)\nonumber \\
 & =\frac{1}{4}\int d^{10}x{\rm Tr}\left[(\lambda^{T}\Gamma^{\mu}C\lambda)(\lambda^{T}\Gamma_{\mu}C\lambda)\right]\int_{0}^{\infty}dt\left(1+\frac{2ae^{-t}}{a^{-1}-e^{-t}a}\right)\,.
\end{align}
However, the last expression includes the divergent integration $\int_{0}^{\infty}dt\:1$,
which can be interpreted as a contribution from the massless part
($L_{0}=0$) in the integration $\frac{1}{L_{0}}=\int_{0}^{\infty}dt\:e^{-tL_{0}}$.
In deriving the equations of motion for the massive part, Eq.~(\ref{eq:EOM_RS}),
we should multiply a projection $\mathcal{P}$, which subtracts the
massless part, on the right-hand side, and this implies that we should
replace $\frac{1}{L_{0}}\to\frac{1}{L_{0}}\mathcal{P}$ in the
above. Namely, inserting the projection, we can subtract the divergent
term as
\begin{align}
\frac{1}{2}\langle\!\langle(\Psi_{0})^{2},\frac{b_{0}}{L_{0}}\mathcal{P}(\Psi_{0})^{2}\rangle\!\rangle & =\frac{1}{4}\int d^{10}x{\rm Tr}\left[(\lambda^{T}\Gamma^{\mu}C\lambda)(\lambda^{T}\Gamma_{\mu}C\lambda)\right]\int_{0}^{\infty}dt\frac{2ae^{-t}}{a^{-1}-e^{-t}a}\nonumber \\
 & =\frac{1}{2}\log\left(2(\sqrt{2}-1)\right)\int d^{10}x{\rm Tr}\left[(\hat{\lambda}^{T}C\Gamma^{\mu}\hat{\lambda})(\hat{\lambda}^{T}C\Gamma_{\mu}\hat{\lambda})\right]\,,
\end{align}
where we have used Eqs.~(\ref{eq:a_def})~and~(\ref{eq:hat_lambda_def}).

Eventually, including a contribution from the massive part to the
massless effective action in both NS and R sectors, Eqs.~(\ref{eq:S_NS_A4})~and~(\ref{eq:S_R_massless}), we have
\begin{align}
S & =\int d^{10}x{\rm \,Tr}\biggl[-\frac{\alpha^{\prime}}{4}\left(\partial_{\mu}A_{\nu}-\partial_{\nu}A_{\mu}-\frac{ig}{\sqrt{2\alpha^{\prime}}}[A_{\mu,}A_{\nu}]\right)^{2}-\frac{\sqrt{\alpha^{\prime}}}{2}i\hat{\lambda}^{T}C\Gamma^{\mu}\left(\partial_{\mu}\hat{\lambda}-\frac{ig}{\sqrt{2\alpha^{\prime}}}[A_{\mu},\hat{\lambda}]\right)\nonumber \\
 & \qquad\qquad\qquad+\frac{1}{2}\log\left(2(\sqrt{2}-1)\right)g^{2}(\hat{\lambda}^{T}C\Gamma^{\mu}\hat{\lambda})(\hat{\lambda}^{T}C\Gamma_{\mu}\hat{\lambda})\biggr]\,.
\end{align}
Rewriting $\sqrt{\alpha^{\prime}}A_{\mu}\to A_{\mu},\ (\alpha^{\prime})^{1/4}\hat{\lambda}\to\hat{\lambda}$,
and $\frac{1}{\sqrt{2}\alpha^{\prime}}g\to g$ for the canonical form,
we obtain
\begin{align}
S & =\int d^{10}x{\rm \,Tr}\biggl[-\frac{1}{4}\left(\partial_{\mu}A_{\nu}-\partial_{\nu}A_{\mu}-ig[A_{\mu,}A_{\nu}]\right)^{2}-\frac{1}{2}i\hat{\lambda}^{T}C\Gamma^{\mu}\left(\partial_{\mu}\hat{\lambda}-ig[A_{\mu},\hat{\lambda}]\right)\nonumber \\
 & \qquad\qquad\qquad+\log\left(2(\sqrt{2}-1)\right)\alpha^{\prime}g^{2}(\hat{\lambda}^{T}C\Gamma^{\mu}\hat{\lambda})(\hat{\lambda}^{T}C\Gamma_{\mu}\hat{\lambda})\biggr]\,,\label{eq:alpha_p_correction}
\end{align}
where the first line is the same as the action of ten-dimensional
SYM and the second line can be regarded as an $\alpha^{\prime}$-correction
due to a superstring.

Concerning the $\alpha^{\prime}$-correction, we comment on the term
of the form $\alpha^{\prime}g{\rm Tr}\left[F_{\mu\nu}F^{\nu\sigma}F_{\sigma}^{\ \mu}\right]$
for the field strength, $F_{\mu\nu}=\partial_{\mu}A_{\nu}-\partial_{\nu}A_{\mu}-ig[A_{\mu,}A_{\nu}]$,
which was investigated in Refs.~\cite{Scherk:1974ca,Tseytlin:1986ti}. In
this section, we have included only the zero-momentum sector for contributions
from the massive part, and then if the above term exists, it should
appear in the form $\alpha^{\prime}g^{4}{\rm Tr}\left[i[A_{\mu},A_{\nu}][A^{\nu},A^{\sigma}][A_{\sigma},A^{\mu}]\right]$,
which is a higher order in $g$ than Eq.~(\ref{eq:alpha_p_correction}).
In this sense, it is necessary to perform further computations including the
nonzero-momentum sector or higher-order terms in $g$ in order to
compare the action from SSFT with the non-Abelian Born--Infeld action
\cite{Tseytlin:1997csa}. 

\section{Induced transformations at the lowest order\label{sec:Induced-transformations}}

Here, we derive the gauge and spacetime supersymmetry transformations
for massless component fields from those of string fields given in
Refs.~\cite{Kunitomo:2015usa,Kunitomo:2016kwh,Erler:2017onq}.
We perform explicit computations up to the lowest order for simplicity.

\subsection{Gauge transformation}

The linearized version of Eqs.~(\ref{eq:A_deltag})~and~(\ref{eq:delta_gPsi})
is given by
\begin{equation}
\delta_{g}^{(0)}\Phi=Q\Lambda+\eta\Omega\,,\qquad\delta_{g}^{(0)}\Psi=Q\lambda\,.
\end{equation}
At the massless level, we have
\begin{equation}
\Lambda_{\varepsilon}=\int\frac{d^{10}k}{(2\pi)^{10}}\varepsilon(k)c\xi\partial\xi e^{-2\phi}e^{ik\cdot X}(0)|0\rangle\,,\qquad\Omega_{\omega}=\int\frac{d^{10}k}{(2\pi)^{10}}\omega(k)\xi e^{ik\cdot X}(0)|0\rangle
\end{equation}
for the gauge transformation parameter string fields in the NS sector,
$\Lambda$ and $\Omega$, and we have no states for $\lambda$-gauge
transformation in the R sector at this level. In order to respect
the partial gauge-fixing condition $\xi_{0}\Phi=0$, we take the gauge
parameter as $\omega(k)=\varepsilon(k)=\varepsilon^{a}(k)T_{a}$ and
we have 
\begin{equation}
Q\Lambda_{\varepsilon}+\eta\Omega_{\varepsilon}=\int\frac{d^{10}k}{(2\pi)^{10}}\left(-\alpha^{\prime}k^{2}\varepsilon(k)\mathcal{V}_{B}(0)-\sqrt{2\alpha^{\prime}}k_{\mu}\varepsilon(k)\mathcal{V}_{A}^{\mu}(0)\right)|0\rangle\,.
\end{equation}
Then, from the linearized gauge transformation at the massless level,
\begin{equation}
\delta_{\varepsilon}^{(0)}\left(\Phi_{A}+\Phi_{B}\right)\equiv Q\Lambda_{\varepsilon}+\eta\Omega_{\varepsilon}\,,
\end{equation}
we have obtained the induced transformation for component fields:
\begin{equation}
\delta_{\varepsilon}^{(0)}B(x)=\alpha^{\prime}\partial^{2}\varepsilon(x)\,,\qquad\delta_{\varepsilon}^{(0)}A_{\mu}(x)=i\sqrt{2\alpha^{\prime}}\partial_{\mu}\varepsilon(x)\,,\qquad\varepsilon(x)=\int\frac{d^{10}k}{(2\pi)^{10}}\varepsilon(k)e^{ik\cdot x}\,.\label{eq:induced_gauge_tr0}
\end{equation}

For the gaugino field $\hat{\lambda}^{\alpha}(x)$, the transformation
is trivial: $\delta_{\varepsilon}^{(0)}\hat{\lambda}^{\alpha}=0$
at the linearized level. In order to get a nontrivial transformation,
we should include the interaction term of SSFT in the gauge transformation.
Expanding Eq.~(\ref{eq:delta_gPsi}) as 
\begin{equation}
\delta_{g(\Lambda)}\Psi=X\eta F\Xi D_{\eta}\left\{ F\Psi,\Lambda\right\} =X\eta\left\{ \Psi,\Lambda\right\} +O(\Phi\Psi)\,,
\end{equation}
we define
\begin{equation}
\delta_{\varepsilon}^{(1)}\Psi_{0}=X\eta\left\{ \Psi_{0},\Lambda_{\varepsilon}\right\} \,.
\end{equation}
Since the star product of $\Psi_{0}$ and $\Lambda_{\varepsilon}$
has $(n_{{\rm gh}},n_{{\rm pic}})=(0,-1/2)$ and its $\phi$-charge
is $-5/2$, we expand as
\begin{equation}
\left\{ \Psi_{0},\Lambda_{\varepsilon}\right\} =\int\frac{d^{10}k}{(2\pi)^{10}}|\varphi^{\dot{\alpha}}(k)\rangle(C^{-1})_{\dot{\alpha}\alpha}\langle\varphi^{{\rm c}\alpha}(-k),\left\{ \Psi_{0},\Lambda_{\varepsilon}\right\} \rangle+\cdots
\end{equation}
at the massless level, where
\begin{equation}
\varphi^{\dot{\alpha}}(k)=c\partial c\,\xi\partial\xi\,S_{(-5/2)}^{\dot{\alpha}}e^{ik\cdot X}(0)|0\rangle\,,\qquad\varphi^{{\rm c}\alpha}(k)=c\eta S_{(1/2)}^{\alpha}e^{ik\cdot X}(0)|0\rangle\,,
\end{equation}
which satisfy the normalization
\begin{equation}
\langle\varphi^{{\rm c}\alpha}(k_{1}),\varphi^{\dot{\alpha}}(k_{2})\rangle=C^{\alpha\dot{\alpha}}(2\pi)^{10}\delta^{10}(k_{1}+k_{2})\,.
\end{equation}
Using the relations
\begin{align}
\langle\varphi^{{\rm c}\alpha}(k_{1}),\left\{ \Psi_{0},\Lambda_{\varepsilon}\right\} \rangle & =\int\frac{d^{10}k_{2}}{(2\pi)^{10}}\int\frac{d^{10}k_{3}}{(2\pi)^{10}}(-1)\left[\lambda_{\dot{\alpha}}(k_{2}),\varepsilon(k_{3})\right]\nonumber \\
 & \qquad\qquad\times C^{\alpha\dot{\alpha}}K^{\alpha'(k_{1}^{2}+k_{2}^{2}+k_{3}^{2})}(2\pi)^{d}\delta^{d}(k_{1}+k_{2}+k_{3})\,,\\
X\eta|\varphi^{\dot{\alpha}}(k)\rangle & =-\mathcal{V}_{\lambda}^{\dot{\alpha}}(k)(0)|0\rangle\,,
\end{align}
we have obtained
\begin{align}
\delta_{\varepsilon}^{(1)}\Psi_{0} & =\int\frac{d^{10}k}{(2\pi)^{10}}\delta_{\varepsilon}^{(1)}\lambda_{\dot{\alpha}}(k)\mathcal{V}_{\lambda}^{\dot{\alpha}}(k)(0)|0\rangle+\cdots\,,\\
\delta_{\varepsilon}^{(1)}\lambda_{\dot{\alpha}}(k) & =\int\frac{d^{10}k_{2}}{(2\pi)^{10}}\int\frac{d^{10}k_{3}}{(2\pi)^{10}}\left[\lambda_{\dot{\alpha}}(k_{2}),\varepsilon(k_{3})\right]K^{\alpha'(k^{2}+k_{2}^{2}+k_{3}^{2})}(2\pi)^{d}\delta^{d}(k_{2}+k_{3}-k)\,,
\end{align}
 and hence the induced gauge transformation of the component field
is
\begin{equation}
\delta_{\varepsilon}^{(1)}\hat{\lambda}^{\alpha}(x)=-\left[\varepsilon(x),\hat{\lambda}^{\alpha}(x)\right]\label{eq:induced_gauge_tr1}
\end{equation}
for small momentum: $K^{\alpha'(k^{2}+k_{2}^{2}+k_{3}^{2})}\sim1$.
Equations~(\ref{eq:induced_gauge_tr0})~and~(\ref{eq:induced_gauge_tr1}) are
consistent with the ordinary gauge transformation of the SYM.

\subsection{Supersymmetry transformation at the linearized level}

First, we derive explicit expressions for equations of motion for
component fields from those for string fields. The equations of motion in Eq.~(\ref{eq:EOM}) are linearized as
\begin{equation}
Q\eta\Phi=0\,,\qquad Q\Psi=0\,.
\end{equation}
At the massless level, we have
\begin{align}
Q\eta\Phi_{0} & =\int\frac{d^{10}k}{(2\pi)^{10}}\Bigl(-2\bigl(B(k)-\sqrt{\frac{\alpha^{\prime}}{2}}k^{\mu}A_{\mu}(k)\bigr)c\eta e^{ik\cdot X}(0)|0\rangle\nonumber \\
 & \qquad\qquad-\sqrt{2\alpha^{\prime}}\bigl(k_{\mu}B(k)-\sqrt{\frac{\alpha^{\prime}}{2}}k^{2}A_{\mu}(k)\bigr)c\partial ce^{-\phi}\psi^{\mu}e^{ik\cdot X}(0)|0\rangle\Bigr)\,,\label{eq:EOM_Phi_0}\\
Q\Psi_{0} & =\int\frac{d^{10}k}{(2\pi)^{10}}\lambda_{\dot{\alpha}}\left(\alpha^{\prime}k^{2}c\partial cS_{(-1/2)}^{\dot{\alpha}}+i\sqrt{\alpha^{\prime}}k_{\mu}(\Gamma^{\mu})_{\ \alpha}^{\dot{\alpha}}\eta cS_{(1/2)}^{\alpha}\right)e^{ik\cdot X}(0)|0\rangle\,,\label{eq:EOM_Psi_0}
\end{align}
which imply that the induced linearized equations of motion are
\begin{equation}
B+i\sqrt{\frac{\alpha^{\prime}}{2}}\partial^{\mu}A_{\mu}=0\,,\qquad i\partial_{\mu}B-\sqrt{\frac{\alpha^{\prime}}{2}}\partial^{2}A_{\mu}=0\,,\qquad\Gamma^{\mu}\partial_{\mu}\hat{\lambda}=0\,,\label{eq:Linearized_EOM_componets}
\end{equation}
in terms of component fields. The first two equations correspond to
the Maxwell equation for the gauge field $A_{\mu}$ and the last equation
corresponds to the Dirac equation, and hence they are consistent with
SYM at the linearized level.\\

The spacetime supersymmetry transformations in Eqs.~(\ref{eq:A_deltaS})~and~(\ref{eq:delta_SPsi}) are linearized as
\begin{equation}
\delta_{\mathcal{S}}^{(0)}\Phi=\mathcal{S}\Xi\Psi\,,\qquad\delta_{\mathcal{S}}^{(0)}\Psi=X\mathcal{S}\eta\Phi\,.\label{eq:delta^0_S}
\end{equation}
At the massless level, the first transformation is computed as
\begin{align}
\delta_{\mathcal{S}}^{(0)}\Phi_{0} & =\mathcal{S}\Xi\Psi_{0}\nonumber \\
 & =\int\frac{d^{10}k}{(2\pi)^{10}}\frac{i}{\sqrt{2}}(\epsilon^{T}\Gamma_{\mu}C\lambda)\mathcal{V}_{A}^{\mu}(k)(0)|0\rangle+\eta\Omega_{0}\,,
\end{align}
where $\Omega_{0}\equiv\xi_{0}\mathcal{S}(\Xi-\xi_{0})\Psi_{0}$ is
a kind of $\Omega$-gauge transformation in Eq.~(\ref{eq:A_deltag}).
Then, up to $\Omega$-gauge transformation, the induced linearized
transformations of bosonic component fields are obtained:
\begin{equation}
\delta_{\mathcal{S}}^{(0)}A_{\mu}=\frac{1}{\sqrt{2}}\bar{\hat{\epsilon}}\Gamma_{\mu}\hat{\lambda}\,,\qquad\delta_{\mathcal{S}}^{(0)}B=0\,,\label{eq:SUSYtr_Amu_B}
\end{equation}
where $\hat{\epsilon}=C\epsilon$ and $\bar{\hat{\epsilon}}=\hat{\epsilon}^{T}C$.
For the second equation of Eq.~(\ref{eq:delta^0_S}), we have calculated
as follows:
\begin{align}
\delta_{\mathcal{S}}^{(0)}\Psi_{0} & =X\mathcal{S}\eta\Phi_{0}\nonumber \\
 & =\int\frac{d^{10}k}{(2\pi)^{10}}\Bigl(\sqrt{\frac{\alpha^{\prime}}{2}}\frac{1}{2}(k_{\mu}A_{\nu}(k)-k_{\nu}A_{\mu}(k))(\epsilon^{T}\Gamma^{\mu\nu})_{\dot{\alpha}}\nonumber \\
 & \qquad\qquad\qquad+\bigl(B(k)+\sqrt{\frac{\alpha^{\prime}}{2}}k^{\mu}A_{\mu}(k)\bigr)\epsilon_{\dot{\alpha}}\Bigr)\mathcal{V}_{\lambda}^{\dot{\alpha}}(k)(0)|0\rangle\,.
\end{align}
This implies that the induced linearized transformation of the fermionic
component field is given by
\begin{equation}
\delta_{\mathcal{S}}^{(0)}\hat{\lambda}=i\sqrt{\frac{\alpha^{\prime}}{2}}\frac{1}{2}(\partial_{\mu}A_{\nu}-\partial_{\nu}A_{\mu})\Gamma^{\mu\nu}\hat{\epsilon}+(B+i\sqrt{\frac{\alpha^{\prime}}{2}}\partial^{\mu}A_{\mu})\hat{\epsilon}\,.\label{eq:SUSYtr_lambda}
\end{equation}
We find that the induced transformations of Eqs.~(\ref{eq:SUSYtr_Amu_B})~and~(\ref{eq:SUSYtr_lambda}) are consistent with the conventional
supersymmetry transformation in the ten-dimensional SYM, up to the
equations of motion in Eq.~(\ref{eq:Linearized_EOM_componets}), at the linearized
level.

\section{Concluding remarks\label{sec:Concluding-remarks}}

In this paper, we have truncated the string fields in both NS and
R sectors in the framework of Kunitomo and Okawa's SSFT up to the lowest
level (massless level) and, by evaluating the action explicitly in
terms of the component fields, we have obtained the ten-dimensional
SYM action plus an extra $O(A_{\mu}^{4})$ term. We have also investigated
a contribution from the massive part in the lowest order in $g$ in the
zero-momentum sector and observed that the extra $O(A_{\mu}^{4})$
cancels in the NS sector and instead an extra $O(\lambda_{\dot{\alpha}}^{4})$
appears from the R sector, which can be interpreted as $\alpha^{\prime}$-correction.\footnote{As noted at the end of Sect.~\ref{sec:Contribution-from-massive},
we should include the nonzero-momentum sector in order to discuss $O(\alpha^{\prime})$-correction
completely.} We have derived the gauge transformation and the spacetime supersymmetry
transformation for the massless component fields induced from those
of string fields at the lowest order. Our explicit calculation implies
that the lowest-level truncation of Kunitomo and Okawa's SSFT action is
consistent with the ten-dimensional SYM theory. 

We have some remaining issues. At the present stage, we have no explicit
formula for the reality condition of the string fields including the R
sector. We expect that the Majorana condition for the massless component
field in the R sector, $\hat{\lambda}^{\dagger}\Gamma^{0}=\hat{\lambda}^{T}C$,
should be imposed by a consistent reality condition for string fields.
It would be interesting to perform similar calculations in other SSFTs such
as $A_{\infty}$-type SSFT and (modified) cubic SSFT. Our concrete
computations in terms of component fields might be useful to find
new methods to construct solutions of the equation of motion such
as SSFT version of the BPS condition.

We hope that our work becomes one of the steps toward physical applications
of SSFT.

\section*{Acknowledgments}

We would like to thank T.~Erler, T.~Kawano, H.~Kunitomo, H.~Nakano,
and T.~Takahashi for valuable discussions and comments. We would also
like to thank the organizers of the conference ``Progress in Quantum
Field Theory and String Theory II'' at Osaka City University and
the 22nd Niigata--Yamagata joint school (YITP-S-17-03) at the National Bandai
Youth Friendship Center, where our work was presented. This work was
supported in part by JSPS Grant-in-Aid for Young Scientists (B) (JP25800134).

\appendix

\section{Convention of the spin fields\label{sec:Convention}}

Here, we summarize our convention for explicit computations including
the R sector, which is based on the method developed in Ref.~\cite{Kostelecky:1986xg}.
The worldsheet fermion $\psi^{\mu}\,(\mu=0,1,\dots,9)$ can be bosonized
using $\phi^{a}\,(a=1,2,\dots,5)$ as
\begin{align}
i\,2^{-1/2}(\psi^{0}\mp\psi^{1}) & \simeq\,e^{\pm\phi^{1}}c_{\pm e_{1}}\,,\label{eq:bosonization1}\\
2^{-1/2}(\psi^{2a-2}\mp i\psi^{2a-1}) & \simeq e^{\pm\phi^{a}}c_{\pm e_{a}}\,,\qquad a=2,3,4,5\,,\label{eq:bosonization2345}
\end{align}
where the operator product expansion (OPE) among $\phi^{a}$ is $\phi^{a}(z)\phi^{b}(w)\sim\delta^{a,b}\log(z-w)\,,\ (a,b=1,\dots,5).$
Involving the bosonized ghost $\phi\equiv\phi^{6}$, such as $\phi^{6}(z)\phi^{6}(w)\sim-\log(z-w)\,,$
the cocycle factor $c_{\lambda\,}(\lambda=\sum_{i=1}^{6}\lambda^{i}e_{i})$
is defined by
\begin{equation}
c_{\lambda}=e^{i\pi\sum_{i,j=1}^{6}\lambda^{i}M_{ij}[{\partial}\phi^{j}]_{0}}\,,\qquad[\partial\phi^{i}]_{0}=\oint\frac{dz}{2\pi i}\,\partial\phi^{i}\,,
\end{equation}
where $M_{ij}\,(i,j=1,2,\dots,6)$ is given by the matrix
\begin{equation}
M=\begin{pmatrix}0 & 0 & 0 & 0 & 0 & 0\\
1 & 0 & 0 & 0 & 0 & 0\\
1 & 1 & 0 & 0 & 0 & 0\\
-1 & 1 & -1 & 0 & 0 & 0\\
1 & 1 & 1 & 1 & 0 & 0\\
-1 & -1 & -1 & -1 & 1 & 0
\end{pmatrix}\,.
\end{equation}
The GSO ($+$) projection can be expressed as
\begin{equation}
\frac{1+(-1)^{G}}{2}\,,\qquad G=\sum_{i=1}^{6}[\partial\phi^{i}]_{0}\,.\label{eq:GSOprojection}
\end{equation}
The spin fields with $n_{\mathrm{pic}}=\pm1/2$ are expressed as
\begin{align}
S_{(\pm1/2)}^{\alpha} & =e^{\sum_{i=1}^{5}A^{i}\phi^{i}\pm\frac{1}{2}\phi}c_{\sum_{i=1}^{5}A^{i}e_{i}\pm\frac{1}{2}e_{6}},\qquad A^{i}=\pm\frac{1}{2},~~~(i=1,\dots,5),\quad\prod_{i=1}^{5}A^{i}>0\,,\\
S_{(\pm1/2)}^{\dot{\alpha}} & =e^{\sum_{i=1}^{5}\dot{A}^{i}\phi^{i}\pm\frac{1}{2}\phi}c_{\sum_{i=1}^{5}\dot{A}^{i}e_{i}\pm\frac{1}{2}e_{6}},\qquad\dot{A}^{i}=\pm\frac{1}{2},~~~(i=1,\dots,5),~~~\prod_{i=1}^{5}\dot{A}^{i}<0\,.
\end{align}
In general, for $S_{\lambda}\equiv e^{\sum_{i=1}^{6}\lambda^{i}\phi^{i}}c_{\lambda}$
with $\lambda=\sum_{i=1}^{6}\lambda^{i}e_{i}$, the OPE is
\begin{equation}
S_{\lambda}(y)S_{\lambda'}(z)=(y-z)^{\lambda^{i}\eta_{ij}\lambda^{\prime j}}e^{\lambda^{i}\phi^{i}(y)+\lambda^{\prime i}\phi^{i}(z)}e^{i\pi\lambda^{i}M_{ij}\lambda^{\prime j}}c_{\lambda+\lambda'}\,,
\end{equation}
where $\eta_{ij}={\rm diag}(1,1,1,1,1,-1)$. Corresponding to the
above convention, we define the $\Gamma$-matrix as
\begin{alignat}{2}
\Gamma^{0} & =-i\sigma_{1}\otimes1_{2}\otimes1_{2}\otimes1_{2}\otimes1_{2}\,, & \qquad\Gamma^{1} & =\sigma_{2}\otimes1_{2}\otimes1_{2}\otimes1_{2}\otimes1_{2}\,,\\
\Gamma^{2} & =\sigma_{3}\otimes\sigma_{2}\otimes1_{2}\otimes1_{2}\otimes1_{2}\,, & \Gamma^{3} & =-\sigma_{3}\otimes\sigma_{1}\otimes1_{2}\otimes1_{2}\otimes1_{2}\,,\\
\Gamma^{4} & =-\sigma_{3}\otimes\sigma_{3}\otimes\sigma_{1}\otimes1_{2}\otimes1_{2}\,, & \Gamma^{5} & =-\sigma_{3}\otimes\sigma_{3}\otimes\sigma_{2}\otimes1_{2}\otimes1_{2}\,,\\
\Gamma^{6} & =-\sigma_{3}\otimes\sigma_{3}\otimes\sigma_{3}\otimes\sigma_{2}\otimes1_{2}\,, & \Gamma^{7} & =\sigma_{3}\otimes\sigma_{3}\otimes\sigma_{3}\otimes\sigma_{1}\otimes1_{2}\,,\\
\Gamma^{8} & =\sigma_{3}\otimes\sigma_{3}\otimes\sigma_{3}\otimes\sigma_{3}\otimes\sigma_{1}\,, & \Gamma^{9} & =\sigma_{3}\otimes\sigma_{3}\otimes\sigma_{3}\otimes\sigma_{3}\otimes\sigma_{2}\,,
\end{alignat}
and $\Gamma^{11}=\Gamma^{0}\Gamma^{1}\Gamma^{2}\Gamma^{3}\Gamma^{4}\Gamma^{5}\Gamma^{6}\Gamma^{7}\Gamma^{8}\Gamma^{9}=\sigma_{3}\otimes\sigma_{3}\otimes\sigma_{3}\otimes\sigma_{3}\otimes\sigma_{3},$
where $\sigma_{i}\,(i=1,2,3)$ is the Pauli matrix, and we take
\begin{align}
C & =e^{\frac{3}{4}\pi i}\sigma_{2}\otimes\sigma_{1}\otimes\sigma_{2}\otimes\sigma_{1}\otimes\sigma_{2}\,,\qquad C^{T}=-C,~~C^{\dagger}=C^{-1}\,.
\end{align}
Then, we have
\begin{align}
\{\Gamma^{\mu},\Gamma^{\nu}\} & =2\eta^{\mu\nu},\qquad C^{-1}\Gamma^{\mu}C=-\Gamma^{\mu T}\,,\qquad(\Gamma^{\mu})^{\dagger}=\Gamma^{0}\Gamma^{\mu}\Gamma^{0}\,.
\end{align}
In the same way as the linear combination of the bosonization, Eqs.~(\ref{eq:bosonization1})~and~(\ref{eq:bosonization2345}), we define
\begin{align}
\Gamma^{\pm e_{1}} & =\frac{i}{\sqrt{2}}(\Gamma^{0}\mp\Gamma^{1})=\sqrt{2}\sigma_{\mp}\otimes1_{2}\otimes1_{2}\otimes1_{2}\otimes1_{2},\\
\Gamma^{\pm e_{2}} & =\frac{1}{\sqrt{2}}(\Gamma^{2}\mp i\Gamma^{3})=\pm\sqrt{2}i\sigma_{3}\otimes\sigma_{\mp}\otimes1_{2}\otimes1_{2}\otimes1_{2},\\
\Gamma^{\pm e_{3}} & =\frac{1}{\sqrt{2}}(\Gamma^{4}\mp i\Gamma^{5})=-\sqrt{2}\sigma_{3}\otimes\sigma_{3}\otimes\sigma_{\mp}\otimes1_{2}\otimes1_{2},\\
\Gamma^{\pm e_{4}} & =\frac{1}{\sqrt{2}}(\Gamma^{6}\mp i\Gamma^{7})=\mp\sqrt{2}i\sigma_{3}\otimes\sigma_{3}\otimes\sigma_{3}\otimes\sigma_{\mp}\otimes1_{2},\\
\Gamma^{\pm e_{5}} & =\frac{1}{\sqrt{2}}(\Gamma^{8}\mp i\Gamma^{9})=\sqrt{2}\sigma_{3}\otimes\sigma_{3}\otimes\sigma_{3}\otimes\sigma_{3}\otimes\sigma_{\mp},
\end{align}
where $\sigma_{+}=\frac{1}{2}(\sigma_{1}+i\sigma_{2}),~\sigma_{-}=\frac{1}{2}(\sigma_{1}-i\sigma_{2}),$
and they can be rewritten as
\begin{equation}
\Gamma^{\pm e_{j}}=(\pm i)^{j-1}\sqrt{2}(\sigma_{3}\otimes)^{j-1}\sigma_{\mp}(\otimes1_{2})^{5-j}\qquad j=1,2,3,4,5.
\end{equation}
Using the above equations, we find
\begin{align}
 & \Gamma^{\pm e_{j}}=\begin{pmatrix}0 & (\Gamma^{\pm e_{j}})_{~\dot{\beta}}^{\alpha}\\
(\Gamma^{\pm e_{j}})_{~\beta}^{\dot{\alpha}} & 0
\end{pmatrix},~~~j=1,2,3,4,5,\\
 & (\Gamma^{\pm e_{j}})_{~\dot{\beta}}^{\alpha}=\delta_{\pm e_{j}+A,\dot{B}}\sqrt{2}\,e^{\pm i\pi\sum_{k=1}^{5}M_{jk}\dot{B}^{k}},\qquad(\Gamma^{\pm e_{j}})_{~\beta}^{\dot{\alpha}}=\delta_{\pm e_{j}+\dot{A},B}\sqrt{2}\,e^{\pm i\pi\sum_{k=1}^{5}M_{jk}B^{k}},\\
 & C=\begin{pmatrix}0 & C^{\alpha\dot{\beta}}\\
C^{\dot{\alpha}\beta} & 0
\end{pmatrix},\\
 & C^{\alpha\dot{\beta}}=\delta_{A+\dot{B},0}\,e^{-i\pi\sum_{i,j=1}^{6}A_{+}^{i}M_{ij}A_{+}^{j}},\qquad A_{+}\equiv(A^{i},1/2),\\
 & C^{\dot{\alpha}\beta}=-\delta_{\dot{A}+B,0}\,e^{-i\pi\sum_{i,j=1}^{6}\dot{A}_{-}^{i}M_{ij}\dot{A}_{-}^{j}},\qquad\dot{A}_{-}\equiv(\dot{A}^{i},-1/2),\\
 & e^{i\pi\sum_{j=1}^{5}M_{6j}A^{j}}=i,\quad e^{i\pi\sum_{j=1}^{5}M_{6j}\dot{A}^{j}}=-i,\quad(\Gamma^{\mu}C)^{\alpha\beta}=(\Gamma^{\mu}C)^{\beta\alpha},\quad(\Gamma^{\mu}C)^{\dot{\alpha}\dot{\beta}}=(\Gamma^{\mu}C)^{\dot{\beta}\dot{\alpha}}\,,
\end{align}
and furthermore, for $C^{-1}=\begin{pmatrix}0 & C_{\alpha\dot{\beta}}^{-1}\\
C_{\dot{\alpha}\beta}^{-1} & 0
\end{pmatrix}$, 
\begin{align}
 & C_{\alpha\dot{\beta}}^{-1}=-\delta_{A+\dot{B},0}\,e^{i\pi\sum_{i,j=1}^{6}A_{+}^{i}M_{ij}A_{+}^{j}},\qquad C_{\dot{\alpha}\beta}^{-1}=\delta_{\dot{A}+B,0}\,e^{i\pi\sum_{i,j=1}^{6}\dot{A}_{-}^{i}M_{ij}\dot{A}_{-}^{j}},\\
 & C^{\alpha\dot{\beta}}C_{\dot{\beta}\gamma}^{-1}=\delta_{\gamma}^{\alpha},\quad C^{\dot{\alpha}\beta}C_{\beta\dot{\gamma}}^{-1}=\delta_{\dot{\gamma}}^{\dot{\alpha}}\,,\quad(C^{-1}\Gamma^{\mu})_{\alpha\beta}=(C^{-1}\Gamma^{\mu})_{\beta\alpha},\quad(C^{-1}\Gamma^{\mu})_{\dot{\alpha}\dot{\beta}}=(C^{-1}\Gamma^{\mu})_{\dot{\beta}\dot{\alpha}}.
\end{align}
Here, we should note that the correspondence of the spinor index is
\begin{eqnarray}
\alpha & \leftrightarrow & A=(\pm1/2,\pm1/2,\pm1/2,\pm1/2,\pm1/2)=\sum_{i=1}^{5}A^{i}e_{i},~~~~\prod_{i=1}^{5}A^{i}>0,\\
\dot{\alpha} & \leftrightarrow & \dot{A}=(\pm1/2,\pm1/2,\pm1/2,\pm1/2,\pm1/2)=\sum_{i=1}^{5}\dot{A}^{i}e_{i},~~~~\prod_{i=1}^{5}\dot{A}^{i}<0,
\end{eqnarray}
for undotted and dotted spinors.

For the spin field with $n_{\mathrm{pic}}=r$,
\begin{equation}
S_{(r)}^{\hat{\alpha}}=e^{\sum_{i=1}^{5}\hat{A}^{i}\phi^{i}+r\phi}c_{\sum_{i=1}^{5}\hat{A}^{i}e_{i}+re_{6}}\,,\qquad\alpha=(\alpha,\dot{\alpha})\ \leftrightarrow\ \hat{A}\,,
\end{equation}
we have the OPE:
\[
\psi^{\mu}(y)S_{(r)}^{\hat{\alpha}}(z)\sim(y-z)^{-\frac{1}{2}}\frac{1}{\sqrt{2}}(\Gamma^{\mu})_{~\hat{\beta}}^{\hat{\alpha}}S_{(r)}^{\hat{\beta}}(z)\,.
\]

\bibliographystyle{utphys}
\bibliography{reference}

\end{document}